\documentclass[runningheads]{llncs}
\usepackage{graphicx}
\usepackage{sgame}
\usepackage{booktabs}
\usepackage{amsmath}

\begin{document}
\title{Evaluating Attacker Risk Behavior in an Internet of Things Ecosystem
\thanks{Funded by the Auerbach Berger Chair in Cybersecurity held by Spiros Mancoridis, at Drexel University}}
%
%
\author{Erick Galinkin\inst{1}\orcidID{0000-0003-1268-9258} \and
John Carter\inst{1}\orcidID{0000-0003-4391-0269} \and
Spiros Mancoridis\inst{1}\orcidID{0000-0001-6354-4281}}
\authorrunning{E. Galinkin et al.}
\institute{Drexel University, Philadelphia PA 19104, USA \\
\email{eg657@drexel.edu}
}
\maketitle             
\begin{abstract}
In cybersecurity, attackers range from brash, unsophisticated script kiddies and cybercriminals to stealthy, patient advanced persistent threats.
When modeling these attackers, we can observe that they demonstrate different risk-seeking and risk-averse behaviors.
This work explores how an attacker's risk seeking or risk averse behavior affects their operations against detection-optimizing defenders in an Internet of Things ecosystem.
Using an evaluation framework which uses real, parametrizable malware, we develop a game that is played by a defender against attackers with a suite of malware that is parameterized to be more aggressive and more stealthy.
These results are evaluated under a framework of exponential utility according to their willingness to accept risk.
We find that against a defender who must choose a single strategy up front, risk-seeking attackers gain more actual utility than risk-averse attackers, particularly in cases where the defender is better equipped than the two attackers anticipate.
Additionally, we empirically confirm that high-risk, high-reward scenarios are more beneficial to risk-seeking attackers like cybercriminals, while low-risk, low-reward scenarios are more beneficial to risk-averse attackers like advanced persistent threats.

\end{abstract}
\section{Introduction}
As a discipline, Cybersecurity has had the privilege of borrowing tools from economics, risk analysis, and even psychology~\cite{anderson2020security}, one of which has been the use of game theory in the context of attack-defense modeling.
However, many of these game theoretic models deal with perfectly rational actors or even an actor who makes no decisions at all, such as a worm -- a self-propagating malware.
Despite the popularity of these models, the majority of real-world attacks are not defenders operating against worms, but rather defenders taking actions against a human attacker.
Our work aims to describe the relationship between attacker strategies and risk-seeking behavior, leverage an Internet of Things (IoT) ecosystem to create and detect actual malware, and analyze the potential effects of varying risk acceptance.

In recent years, smart home and smart office devices have become more widely available -- smart locks, smart thermostats, smart fridges, and even smart oven ranges have cropped up.
Consequently, the Internet of Things has become a fresh battleground for security.
Malware like the Mirai botnet~\cite{antonakakis2017understanding} has turned thousands of largely insecure, simple devices into a widely distributed, mass of unwitting soldiers.
To that end, considerable work has been done to attempt to secure IoT systems, such as those in healthcare~\cite{abie2012risk}, where adaptive approaches have proven useful.

We note however, that definitions of IoT vary widely, as IoT are typically associated with consumer-grade products, and correspondingly are often overlooked by corporate information technology and security teams~\cite{heiland2019iot}.
This lack of clarity makes it quite difficult for defenders to know what the best defensive technologies are, and in cases where any security technology is deployed, often results in an approach to IoT devices that fails to mitigate the entirety of the threat.
This ultimately leads to an environment ripe for attackers to exploit.

In security game theoretic literature, attacker-specific strategies and decision making remains an under explored element of security game and decision theory that often necessitates significant amounts of uncertainty modeling~\cite{Chatterjee2015QuantifyingPayoffs}.
Many factors go into an attacker's decision making: the available exploits and payloads, the systems in the environment that are vulnerable to their exploits, what attackers know about the defender's strategy, and risk-seeking or risk-averseness of attackers.
Risk aversion and incentives have been studied extensively since the seminal work of Holt and Laury~\cite{holt2002risk}, showing the incentive effects of different payoffs. 
This is most often used in financial modeling, showing where risk seeking bidders fare well versus risk averse bidders, and establishing strategies based on portfolio simulations. 
We aim to apply a similar framework to one facet of security games, enabling more thoughtful modeling of attacker behavior.

\section{Background}
Though our data is collected from an IoT ecosystem, our primary focus is on attacker strategies.
As a result, the works most closely related to ours are two papers by Chatterjee \textit{et al.}~\cite{Chatterjee2015QuantifyingPayoffs,Chatterjee2016PropagatingAnalysis}.
In their work on quantifying attacker payoffs, Chatterjee \textit{et al.} leverage a leader-follower security game to compute a range of possible payoffs for an attacker.
This builds off of the attacker response function developed by Kiekintveld \textit{et al.}~\cite{kiekintveld2011approximation}, that generates probabilities of attack for various targets.

Chatterjee \textit{et al.}'s other work on propagating uncertainties evaluates the uncertainties in attacker payoffs and leverages Monte Carlo sampling and bound analysis to estimate attacker payoffs in partially observable security games.
These works acknowledge the problem of describing and dealing with the large number of potential attackers and our inability to know what exploits and payloads are available to those attackers.
Our work veers away from this probabilistic approach and deals instead with data generated by real malware in a real IoT ecosystem. 

\begin{figure}
    \centering
    \includegraphics[width=0.8\textwidth]{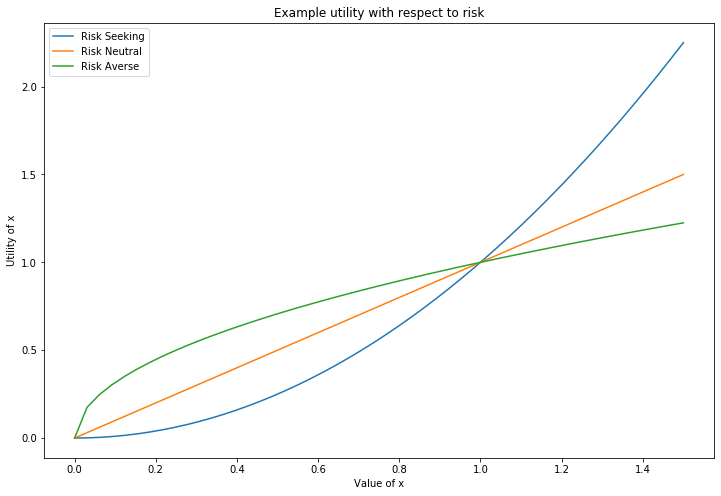}
    \caption{Example plot of risk-seeking, risk-averse, and risk-neutral utility, $u(x)$, derived from a good of value $x$}
    \label{fig:examplerisk}
\end{figure}

We define risk aversion as a utility function that is concave and strictly increasing. 
Risk seeking, then, is a utility function which is convex and strictly increasing.
Example plots of risk-seeking, risk-averse, and risk-neutral utility are shown in Figure~\ref{fig:examplerisk}.
This relationship to risk has been used in behavioral game theory~\cite{camerer2010behavioural} to describe investor behavior.
From a security perspective, this corresponds to the dichotomy between cybercriminals -- threat actors who seek to quickly maximize their profits and engage in risk-seeking behavior, and advanced persistent threats -- threat actors who are willing to wait long periods and want to minimize their chances for detection.

Risk aversion has also been tied to uncertainty aversion~\cite{calford2020uncertainty}, building on early experiments~\cite{camerer1994ambiguity} in the subjective beliefs of actors in an environment.
The expected utility model does not fully capture uncertainty aversion, but this is a first step in the space.
In the context of security games, attackers and defenders both leverage their beliefs about the capabilities of their adversary, but there is significant uncertainty about the interplay between attacker capabilities and defender capabilities.
Kiekintveld \textit{et al.}~\cite{kiekintveld2011approximation} acknowledge the significant challenges faced about both attacker capabilities and the efficacy of defender responses against attacks.
Our work leverages the idea that human operators will choose their tools not only based on what is available to them and what is likely to help them accomplish their objectives, but also their perception of value relative to the probability of success.

\section{Methods} \label{sec:ecosystem}
Leveraging a simple IoT ecosystem with parameterizable malware living on an infected router, we compute the probability of detection for each detection strategy and each malware family with default parameters.
Our ecosystem is comprised of an IP camera, an infected router, and an attacker laptop which is exfiltrating data from the infected router. 
In our ecosystem, there are three different families of malware: ransomware, keylogger, and cryptominer.
Each of these malware has customizable parameters that control the rate of their operation and the size of data that is sent to the malware command and control.
By default, the ransomware will communicate with its command and control server every 15 seconds; the keylogger will exfiltrate 2 keypresses every 0.1 seconds; and the cryptominer will communicate with its command and control server every 0.1 seconds.
These rates can be adjusted up and down or modified to change at each communication interval to create an unlimited number of variants.

\begin{figure}
    \centering
    \includegraphics[width=\textwidth]{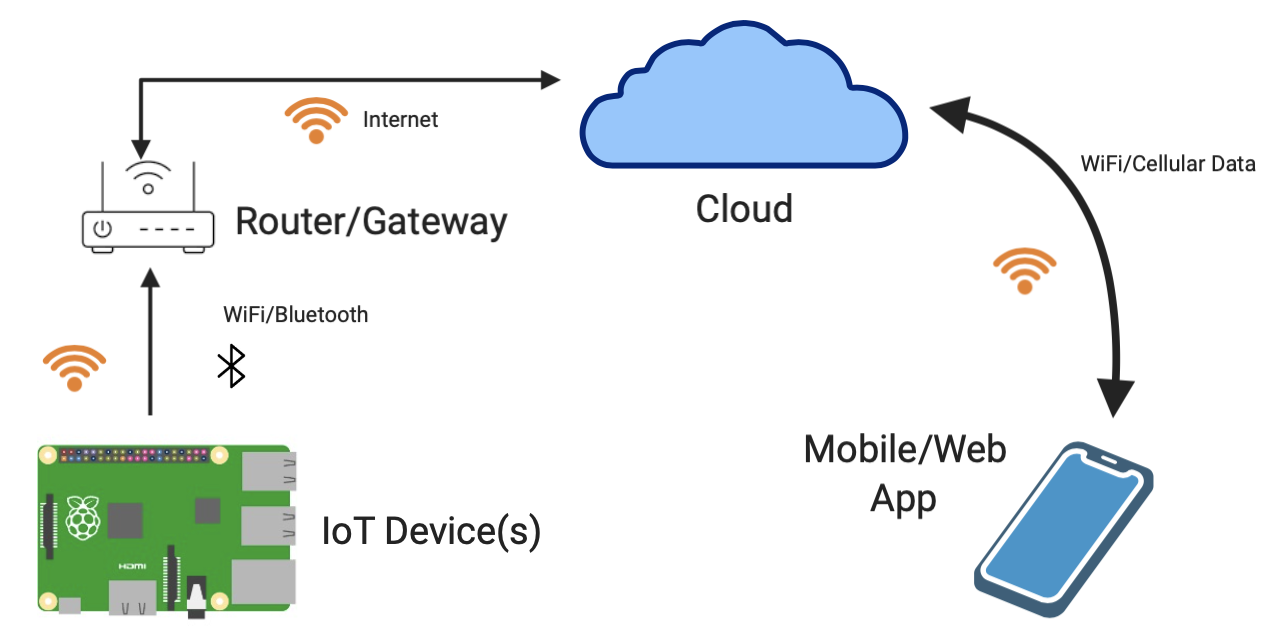}
    \caption{An Example IoT Ecosystem}
    \label{fig:ecosystem}
\end{figure}

Detection strategies consist of three support vector machine (SVM) malware detectors.
These three SVMs differ in terms of the data that they leverage to conclude whether or not malware is present.
One leverages system call (syscall) data from the endpoint and a bag-of-$n$-grams approach to create feature vectors; the second leverages features from network packets; and the third merges the syscall data and the packet data.
We refer to these three detectors by the name of the data that they use: syscall, packets, and merged.

In order to establish a baseline for the various defensive capabilities, we generate 15 malware samples per malware family using the default parameters.
Each sample is run over a 7-minute window in which each of the three detection strategies are able to observe the environment's behavior and estimate a probability that malicious activity is present in the ecosystem. 
If the malware detector crosses the threshold for maliciousness, it sets an alert and the malware is considered caught.
If the detector does not cross the threshold, the malware is not detected and is considered to have successfully accomplished its task.
We average these detection rates over our 15 trials per combination of malware and detector to compute a probability of detection for each detector.

These baseline probabilities for each detection strategy to find each family of malware establish a detection probability matrix that allows defenders to select the strategy with the highest probability of detecting malicious activity in general.
Having committed to their strategy, we simulate attackers with different willingness to accept risk based on their expected payoffs for their malware.
These attackers will target our defender, and the efficacy of their attacks informs our analysis in Section~\ref{sec:eval}.

\section{Defining the Game} \label{sec:game}
Our game proceeds in two stages. 
First, the defender selects their optimal strategy based upon the average-case probability matrix in Table~\ref{tab:prob}, where the malware family names have been abbreviated for readability.
In our game, this probability matrix is common knowledge for both players -- in the real-world, this is analagous to advertised detection rates by different commercial security solutions.

Defenders are often budget constrained, so attackers cannot know what defensive strategy a defender will choose but can be confident that once a defender's strategy is chosen, it is fixed in the short-term. 
In general, collecting packets is low-cost both in terms of both deployment and maintenance costs. 
Collecting syscalls involves deploying agents to endpoints and licensing software for each endpoint, a substantial cost to many businesses.
The merged data uses both syscall and packet data, making it the most expensive solution to deploy.
We assume that our defender is not constrained by resources and chooses the detection strategy with the best average detection rates -- the system with the merged detections, which has an average detection rate across all 3 malware families of 98.01\%.

\begin{table}[h]
\hspace*{\fill}%
\begin{game}{3}{3}[Attacker][Defender]
            &   Syscall     &   Packets     &   Merged  \\
Keylog      &   96.53\%     &   88.76\%     &   96.35\% \\
Mine        &   96.14\%     &   96.54\%     &   97.76\% \\
Ransom      &   99.92\%     &   99.38\%     &   99.91\% 
\end{game} \hspace*{\fill}%
\caption{Probability of attacker detection} \label{tab:prob}
\end{table}

In this case, an attacker will consider three different malware families: keyloggers, cryptominers, and ransomware.
Our three families of malware have different potential payoffs for successful attackers.
The keylogger collects keystrokes and may collect usernames and passwords, or other sensitive data which can be sold.
Overall, the keylogger is of middling value.
A cryptominer can remain resident on a system for quite some time, mining cryptocurrency like bitcoin.
However, it may require a significant amount of time for a block to be mined and for the miner to yield a payoff, so it has a low value.
Ransomware hits quite quickly and offers a significant payoff -- assuming that a victim pays -- giving it the highest value for successful attacks.
Thus, for successful attacks:
\[v_{ransomware} \succ v_{keylogger} \succ v_{cryptominer}\]

We assume that all attackers are seeking to maximize their utility over their set of attacks.
Each attacker has a utility function $u(v_i)$, which asserts the utility they gain from the value of a successful attack, defined as an attack which goes unnoticed.
Their expected utility for attack $i$ is then $(1 - p_i) \cdot u(v_i)$, where $p_i$ is the average probability of detection for the attack from Table~\ref{tab:prob}.
Given the detection rates, the probability of detecting cryptominers is higher than that of detecting keyloggers, and keylogger utility is greater than that of cryptominers.
Therefore, no rational attacker would choose the cryptominer, and so the remainder of our analysis concerns only ransomware and keyloggers.

We leverage exponential utility for the attackers, defining our $u(v_i)$:
\begin{equation}
    u(v_i) =
    \begin{cases}
    \frac{1-e^{-\alpha v_i}}{\alpha} & \alpha \neq 0 \\ \label{eqn:utility}
    v_i & \alpha = 0
    \end{cases}
\end{equation}
Exponential utility has been well studied in behavioral economics~\cite{camerer2004advances} and implies constant absolute risk aversion (CARA) with a risk aversion coefficient equal to the constant $\alpha$ such that for valuation $v_i$:
\[\alpha = \frac{-u''(v_i)}{u'(v_i)}\]
Where $u'(v_i)$ and $u''(v_i)$ are the first and second derivatives of the utility function, respectively.
This provides a parameter $\alpha$ such that $\alpha = 0$ for risk-neutral attackers, $\alpha > 0$ for risk-averse attackers, and $\alpha < 0$ for risk-seeking attackers.

\begin{figure}[h!]
    \centering
    \includegraphics[width=\textwidth]{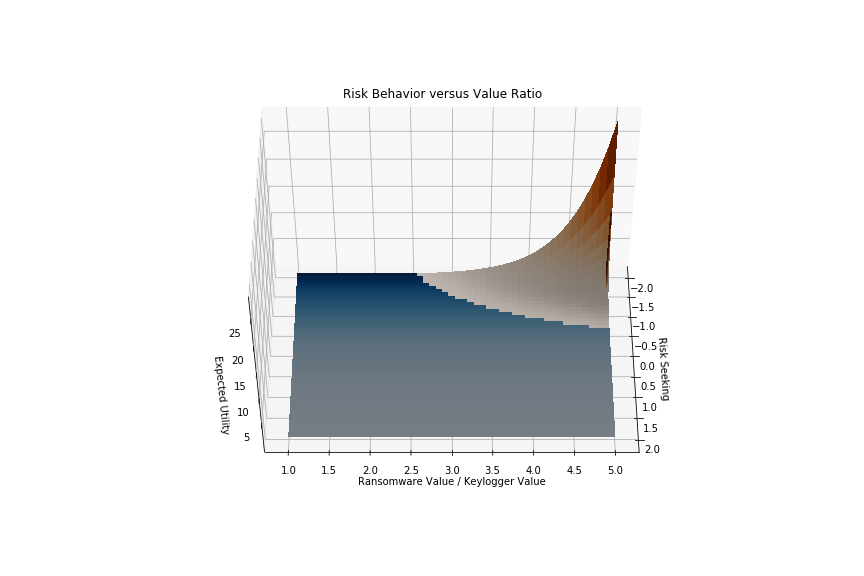}
    \caption{3d plot showing the relationship between risk seeking behavior and the relative value of ransomware over a keylogger}
    \label{fig:value_behavior}
\end{figure}

The factor that will determine the strategy of the attacker then, is a function of the risk behavior of the attacker and the value ratio between our two rational malware choices -- keyloggers and ransomware.
Figure~\ref{fig:value_behavior} shows that there is a relationship between the relative value of the ransomware versus the keylogger and the risk seeking behavior of the attacker.
The blue-shaded areas are the regions where a keylogger is preferential, while the orange-shaded areas are the regions where ransomware is preferable. 

We conclude that given the higher value of ransomware coupled with a higher probability of detection, an attacker's propensity to use ransomware or a keylogger is directly tied to their constant of risk aversion. 
Specifically, more negative CARA corresponds to a greater propensity to use ransomware and more positive CARA corresponds to a greater propensity to use keyloggers.
Based on the data collected from our environment, risk-neutral attackers with $\alpha = 0$ will tend to use keyloggers.

\begin{figure}[h!]
    \centering
    \includegraphics[width=\textwidth]{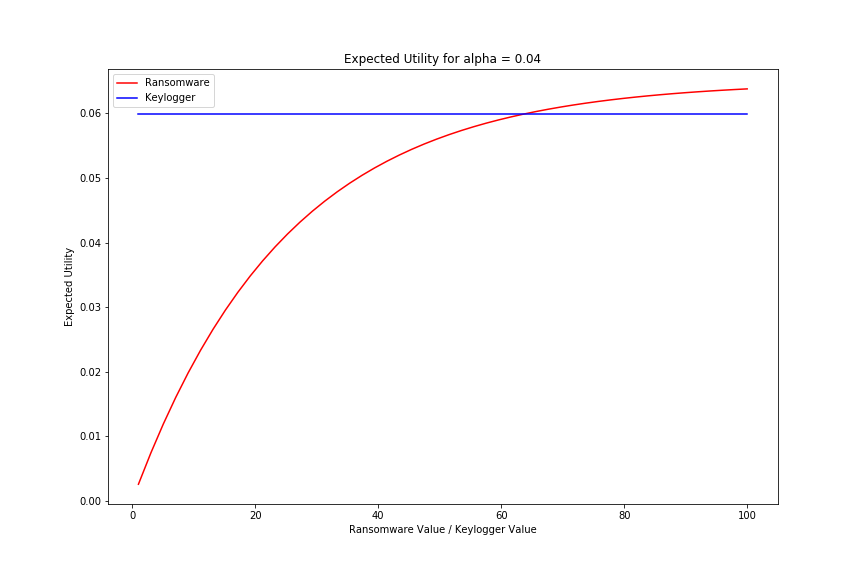}
    \caption{Expected utility for $\alpha = 0.04$}
    \label{fig:eu04}
\end{figure}

Empirically, a rational, risk-averse attacker will still choose ransomware in the case where their expected utility is higher than that of a keylogger.
As an example, let the attacker be risk-averse with $\alpha = 0.04$ and let the keylogger valuation $v_k = 1$.
Using the detection probabilities in Table~\ref{tab:prob}, we find that the expected utility for $v_k$ is 0.059992158.
This means that $u(v_r) > u(v_k)$ only for $v_r \geq 64.08456 v_k$, which yields expected utility 0.059992159 as we can observe in Figure~\ref{fig:eu04}.

Therefore, the risk-averse attacker will choose ransomware only when the value of ransomware, $v_r$ is more than 64 times as large.
These conclusions assume that these malware values and detection rates are static.
However, the customization capabilities of the malware ecosystem allows a risk-seeking attacker to attempt to increase the value of their malware variants and allows a risk-averse attacker to attempt to decrease the detection rates of their malware variants.
In this case, the attacker may manipulate the parameters in such a way that they believe they can improve their expected utility.

\section{Evaluation} \label{sec:eval}
The defender aims to minimize the threat by choosing the strategy with the highest probability of stopping an arbitrary attack and so chooses the merged detection system.
Our attackers, not knowing the malware parameters used to generate the probabilities in Table~\ref{tab:prob}, may choose their own parameters for the malware.
As an illustrative example, we define a risk-seeking attacker with $\alpha = -0.04$, and a risk-averse attacker with $\alpha = 0.04$, who parameterize a suite of malware to be more aggressive and more stealthy as follows:
\begin{enumerate}
    \item Aggressive ransomware: Exfiltration every 2 seconds
    \item Aggressive keylogger: Exfiltration every 0.05 seconds
    \item Stealthy ransomware: Exfiltration every 45 seconds
    \item Stealthy keylogger: Exfiltration every 2 seconds
\end{enumerate}

These malware parameters can be viewed as impacting the value of successful attack, $v_i$. 
A more aggressive ransomware variant will complete encryption more quickly, reducing the time a defender can take to interrupt the operation.
Meanwhile, a more aggressive keylogger variant is likely to exfiltrate greater amounts of data, leading to more potential income.
The stealthier variants space out exfiltration, a technique which is commonly assumed to result in lower detection rates.
These variants are evaluated against the defender's chosen detection strategy, with the common expectation being that a risk-seeking attacker will find more value in aggressive variants while a risk-averse attacker would favor stealthy variants.

\begin{table}
\centering
\begin{tabular}{@{}l|cc@{}}
\toprule
                 & \textbf{Ransomware}     & \textbf{Keylogger} \\ \hline
\textbf{Aggressive}    & 99.958\%           & 100\% \\
\textbf{Stealthy}      & 99.956\%           & 99.72\%  \\ \hline
\bottomrule
\end{tabular}
\caption{Detection results for risk-seeking and risk-averse malware variant} \label{tab:results}
\end{table}

The variants described above are detected at the rates depicted in Table~\ref{tab:results}.
We find that these detection rates are higher than expected in the average case across variants.
Given the actual rates of detection, we compute the attacker's actual derived utility by computing:
\[(1 - p) u(x)\]
Where $p$ is the actual rate of detection and $u(x)$ is the same utility function defined by Equation~\ref{eqn:utility}.
We set $x = 1$ for default parameters on each malware variant to compare results within a malware family.
To establish the value of aggressive and stealthy variants, we use the ratio of the default and modified exfiltration rate to establish the value.
For readability, the utility values shown are scaled by a factor of $1000$.

\begin{table}
\centering
\begin{tabular}{@{}l|ccc|ccc@{}}
\toprule
                         & \multicolumn{3}{c}{\textbf{Ransomware}}   & \multicolumn{3}{c}{\textbf{Keylogger}} \\ \hline
                         & \textbf{Aggressive}  & \textbf{Stealthy}   & \textbf{Expected}      & \textbf{Aggressive}  & \textbf{Stealthy}   & \textbf{Expected} \\ \hline
\textbf{Risk-Seeking}    & 3.674                & 0.148               & 2.653                  & 0.0                  & 0.140               & 62.440 \\
\textbf{Risk-Averse}     & 2.721                & 0.146               & 2.549                  & 0.0                  & 0.140               & 59.992 \\ \hline
\bottomrule
\end{tabular}
\caption{Aggressive, stealthy, and expected utilities for attackers across malware variants, scaled by 1000} \label{tab:utility}
\end{table}

We observe from the results in Table~\ref{tab:utility} that across malware families, risk-averse attackers gain lower nominal actual utility in all cases -- a function of the very high probability of being detected.
Although the aggressive ransomware had the highest detection rate, it conferred the most actual utility relative to expectations for both attacker types.
This is largely due to the high value it received, $v = 15/2$ compared with the stealthy variant $v = 15/45$ and the stealthy keylogger $v = 0.1/2$ and the fact that all detection rates exceeded their expected rate.

A rational actor will choose whatever option \textit{a priori} offers the highest expected value. 
By assumption in Section~\ref{sec:game}, the value of ransomware is greater than that of keyloggers by some unknown amount. 
To incentivize ransomware use, this value must exceed 64 times the value of a keylogger for risk-averse attackers and exceed 16 times the value of a keylogger for risk-seeking attackers given the detection rates in Table~\ref{tab:prob}.
This fact stems from three dimensional analysis of the value ratio and risk sensitivity, visualized in Figure~\ref{fig:value_behavior}.

On observation, the actual utility of the aggressive ransomware variant is 26.24 times the actual utility of the stealthy keylogger for risk-seeking attackers and 19.44 times the actual utility of the stealthy keylogger for risk-averse attackers.
If attackers used either of these values \textit{a priori} to compute the expected value of ransomware, the risk-seeking attacker would choose ransomware, but the risk-averse attacker would choose a keylogger.
In this case, the risk-seeking attacker would be satisfied with their outcome relative to expectations, while the risk-averse attacker would be unsatisfied relative to expectations.

\section{Conclusion}
This work has explored how under the framework of exponential utility, a well-studied case of expected utility theory satisfying CARA, attackers who are operating against an unknown defender will choose their attack strategy based on the combination of their risk appetite and the expected value of the available strategies.
We conclude that in environments where detection rates are higher than expected, risk-seeking attackers are more satisfied with their \textit{a priori} best choice, while risk-averse attackers are less satisfied.
This suggests that risk-averse attackers may seek environments that are more vulnerable and less well-guarded, while risk-seeking attackers are less swayed by a defender's strategy.

Crucially, our work makes a step forward in empirically evaluating attacker strategies using real, parameterizable malware families.
This has long been a difficult task, as valuations and utility functions are highly subjective.
This work demonstrates that given a fairly small set of assumptions, we can test expected utility models and their efficacy.
In future work, we will explore the intricate relationships between attacker risk attitudes and past successes, leveraging a Bayesian model to set the expected value for a successful attack.

From a defender's perspective, we would like to deal directly with the different risk attitudes toward gains and losses.
Different defensive strategies have different costs and different attacks can result in more or less damage to systems.
Using Tversky and Kahneman's cumulative prospect theory~\cite{tversky1992advances}, we can frame the risk of loss differently from that of retaining existing resources or gaining more capital.
Using a cumulative prospect theory model would also allow for better generalization to unknown attacks, since we can consider arbitrary outcome distributions.

Our defensive parameters and malware parameters are also highly configurable and have a continuous range, allowing for unlimited configurations of both offensive and defensive strategies.
The specific parameterization will alter both our detection rates and valuations in ways which are difficult to predict.
In future work, we seek to expand on the limited number of offensive and defensive strategies considered in this paper and instead seek to understand how these parameters can be chosen as a function of the risk appetites of both attackers and defenders.
We will do this by finding optimal parameter settings for both attackers and defenders and developing a significant way of directly relating parameterization with value, using the framework of Dempster-Shafer theory~\cite{shafer1992dempster} to better capture the uncertainty of the players.
This extension of our current work would help frame the relationship between attackers and defenders and their risk appetites in a way that captures the overwhelming number of malware variants and defensive architectures that exist in the real world.

\bibliographystyle{splncs04}
\bibliography{references}

\end{document}